\documentclass{PoS}

\def\mtop{m_{\rm{top}}}

\def\mtoppole{m_{\rm{top}}^{\rm{pole}}}

\def\rtt{R_{\rm{3/2}}}

\def\ttbar{t\bar{t}}

\def\mlb{m_{\ell b}^{\rm{reco}}}

\def\sigttbar{\sigma_{\ttbar}}

\def\GeV{\rm{GeV}}
\def\TeV{\rm{TeV}}

\def\fbm1{\rm fb^{-1}}
\def\stat{\rm stat.}
\def\syst{\rm syst.}

\title{Measurements of the top-quark mass with the ATLAS detector}

\ShortTitle{Measurements of the top-quark mass with the ATLAS detector}

\author{\speaker{Teresa Barillari}\\
             On behalf of the ATLAS Collaboration \\
             Max-Planck-Institute f{\"u}r Physik\\
             F{\"o}hringer Ring 6, 80805 Muenchen, Germany \\
             E-mail: \email{barilla@mppmu.mpg.de}}

\abstract{The top-quark mass is one of the fundamental parameters of the
  Standard Model of particle physics.
  The latest ATLAS measurements of the top-quark
  mass in top quark pair final states are presented. Measurements
  use di-lepton, lepton+jets and all-jets final states and their
  combination is performed. Measurements of the top-quark pole mass
  based on precision theoretical QCD calculations for lepton
  kinematic distribution and for top quark pair production with an
  additional jet are also presented.
  }

\FullConference{ICHEP 2018, XXXIX International Conference
                 on High Energy Physics\\
		4-11 July 2018\\
		Seoul, South Korea}

\begin{document}

\section{Introduction}
\label{sec:intro}

The top-quark mass ($\mtop$) is a fundamental parameter of the 
Standard Model (SM) and its precise value is indispensable for 
predictions of cross sections at the LHC. After the Higgs boson 
discovery at the LHC~\cite{Aad:2012tfa,Chatrchyan:2012ufa} and in the 
current absence of direct evidence for new physics beyond the SM, 
precision theory predictions confronted with precision measurements 
are becoming an important area of research for self-consistency tests 
of the SM and in searching for new physics 
phenomena~\cite{Degrassi:2012ry}.
The $\mtop$ measurements proceed then via kinematic reconstruction of the
top quark's decay products, a $W$ boson and a $b$-quark jet, $b$-jet, and
comparisons to Monte Carlo (MC) simulations are done. 
These $\mtop$ measurements are often referred to as MC top-quark mass 
($\mtop^{\rm MC}$) measurements.
In many Quantum Chromodynamics (QCD) calculations the top-quark 
pole mass ($\mtoppole$), corresponding to the definition of the mass 
of a free particle, is used as the conventional scheme choice. 
Recent studies estimate the value of $\mtop^{\rm MC}$ differs from 
$\mtoppole$ by ${\cal{O}}(1\,\GeV$)~\cite{Moch:2014tta}.
However, more recently it has been quoted that this difference could 
be much smaller~\cite{Nason:2017cxd}.
The $\mtoppole$ can be measured from inclusive $\ttbar$ production 
cross section ($\sigttbar$).
However, this $\mtoppole$ determination is currently less precise 
than the achieved $\mtop^{\rm MC}$ measurements. 
This is due to the weak sensitivity of the inclusive
$\sigttbar$ to the $\mtoppole$, but also to the large uncertainties 
on the factorisation and renormalisation scales and the proton parton
distribution function (PDF).
In the following the latest ATLAS results on $\mtop^{\rm MC}$, or 
just $\mtop$, measurements in the di-lepton~\cite{Aaboud:2016igd}, 
the lepton+jets~\cite{Aaboud:2018zbu}, and in the all-jets~\cite{Aaboud:2017mae}
$\ttbar$ decay channel using data at a centre of mass energy 
$\sqrt{s} = 8\,\TeV$ for an integrated luminosity of $20.2\,\fbm1$
are presented, see Section~\ref{sec:mtop}. 
The $\mtoppole$ value is determined  from inclusive $\sigttbar$ 
measurements in the di-lepton $\ttbar$ decay channel~\cite{Aaboud:2017ujq}
and from normalised differential cross section measurements for events 
with $\ttbar$ in association with at least one jet, $\ttbar+1$-jet~\cite{Aad:2015waa}, see Section~\ref{sec:polemass}.
These analyses use data collected at $\sqrt{s} = 8\,\TeV$ and $7\,\TeV$
with an integrated luminosity of $20.2\,\fbm1$ and $4.6\,\fbm1$
respectively.

\section{Measurements of the top-quark mass at $\sqrt{s}= 8\,\GeV$}
\label{sec:mtop}

\begin{figure}[htp]
  \centering
       {\resizebox{0.31\textwidth}{!}{%
        \includegraphics{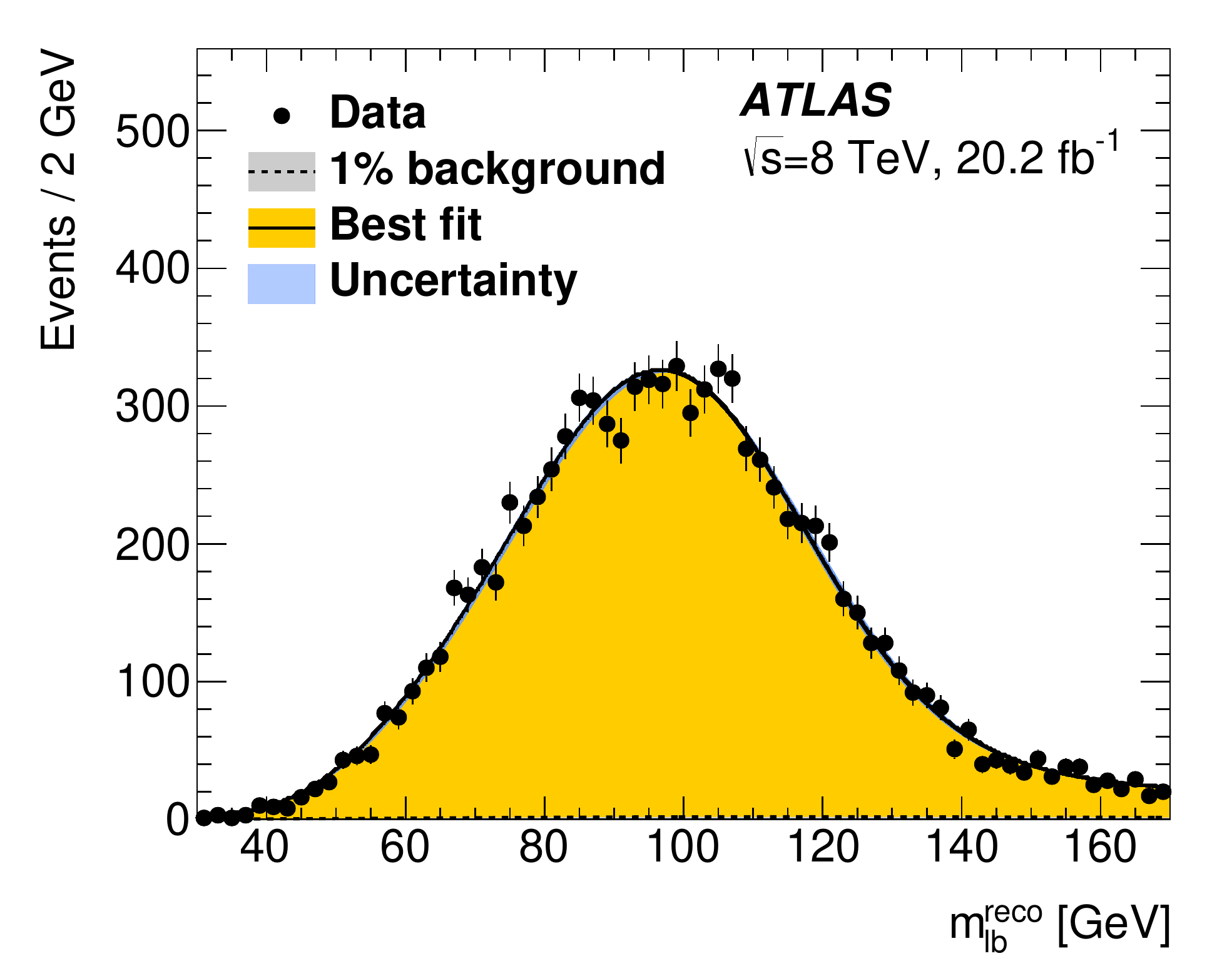}}}
        {\resizebox{0.34\textwidth}{!}{%
        \includegraphics{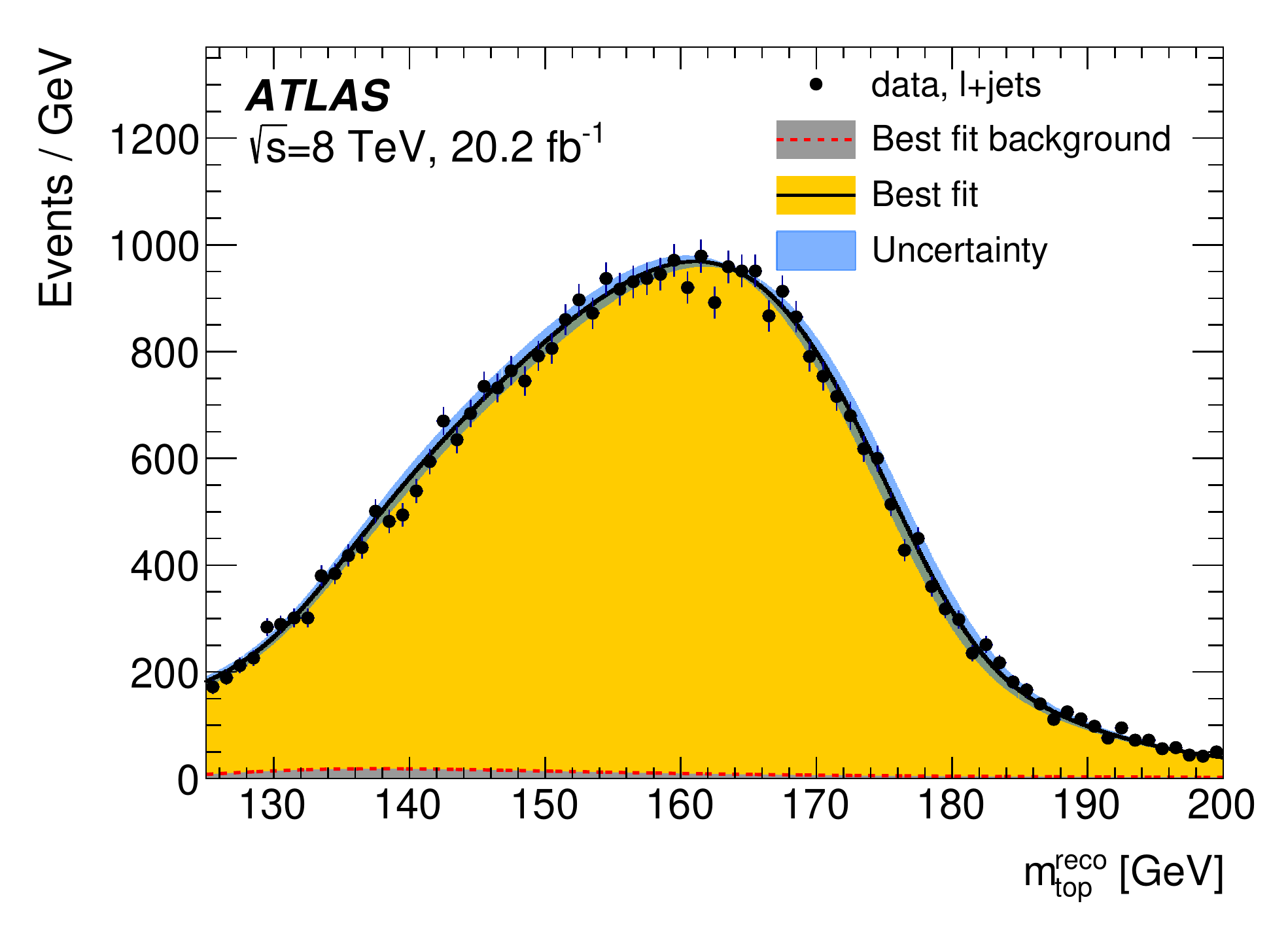}}}
        {\resizebox{0.28\textwidth}{!}{%
        \includegraphics{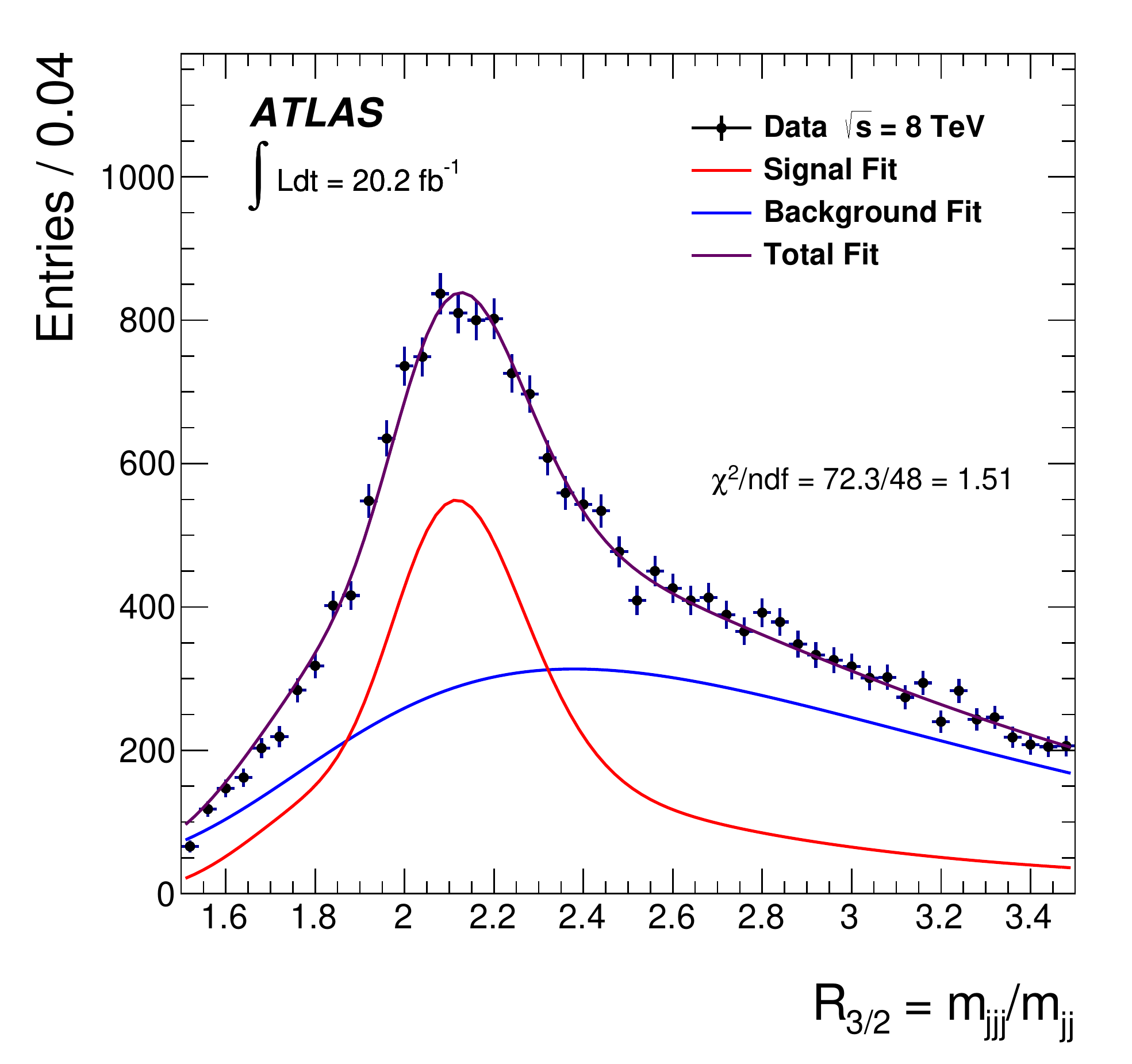}}}
  \caption{
    The left plot, for the $\ttbar$ di-lepton analysis~\cite{Aaboud:2016igd},
    and the central plot, for the
    $\ttbar$ lepton+jets analysis~\cite{Aaboud:2018zbu},
     show the distribution 
     for data together with the fitted 
     probability density functions for the background alone, barely 
     visible at the bottom of the figure, 
     and for the sum of signal and background. Right plot, all-jets
     analysis~\cite{Aaboud:2017mae}, shows
     the $\rtt$ distribution in data with the
     total fit (magenta) and its decomposition into signal (red)
     and the multi-jet background (blue). All errors shown are
     statistical only.
  }
  \label{fig:dilepton}
\end{figure}

A measurement of $\mtop$ is obtained in the $\ttbar\to$ di-lepton decay 
channel using data at $\sqrt{s} = 8\, \TeV$~\cite{Aaboud:2016igd}. 
The analysis exploits the decay 
$\ttbar\to W^+ b W^- \bar{b}\to \ell^{+}\ell^{-}\nu\bar{\nu}b\bar{b}$, 
where both $W$ bosons decay into a charged lepton and its corresponding 
neutrino.  
The event selection described in Ref.~\cite{Aaboud:2016igd}
was optimised to achieve the smallest total uncertainty.  
In the analysis, $b$-jets are selected, and two leptons are taken as 
the leptons from the $W$ decays.
From the two possible assignments of the two 
pairs, the combination leading to the lowest average invariant mass 
of the two lepton-$b$-jet pairs ($\mlb$) is retained.
To perform the template parameterisation, see Ref.~\cite{Aaboud:2016igd}, 
the reconstructed $\mlb$ quantity is used.
The resulting template fit function based on simulated distributions 
of $\mlb$ has $\mtop$ as the only free parameter and an unbinned 
likelihood maximisation gives a value 
of $\mtop = 172.99\,\pm\,0.41\, (\stat)\,\pm\,0.72\,(\syst)\,\GeV$. 
Figure~\ref{fig:dilepton}, left plot, shows the distribution 
obtained with data together with the fitted probability density 
functions for the background (hardly visible at the 
bottom of the figure).
This $\mtop$ result is
the most precise single result in this decay channel to date. 
The biggest systematic uncertainties comes from the measurement
of the jet energies, through the jet energy scale (JES) and relative
$b$-to-light-JES ($b$JES).
The next most recent measurement of $\mtop$ in the $\ttbar\to$ lepton+jets
decay channel performed in ATLAS~\cite{Aaboud:2018zbu}, looks for events with
the decay $\ttbar\to W^+ b W^- \bar{b}\to q\bar{q}'b \ell \nu \bar{b}$.
Here one $W$ boson decays into a charged lepton ($\ell$ is 
$e$ or $\mu$ including $\tau\to e$, $\mu$ decays) and
a neutrino ($\nu$), and the other into a pair of quarks.
A full kinematic reconstruction of the event is done with a likelihood fit
algorithm.
For the measurement of $\mtop$, the event selection is
optimised~\cite{Aaboud:2018zbu} by using a multivariate Boosted Decision
Tree, BDT, algorithm. 
The analysis uses then a three-dimensional template fit technique 
which determines $\mtop$ together with the jet energy scale factors (JSF)
and relative $b$-to-light-JSF ($b$JSF).
Figure~\ref{fig:dilepton}, central plot, shows the the $\mtop$ distribution 
in the data with statistical uncertainties together with the corresponding
fitted probability density functions for the background alone and for the
sum of signal and background.
The achieved measured value of $\mtop$ obtained using the BDT selection 
is $\mtop = 172.08\,\pm\,0.38\,(\stat)\,\pm \,0.82\,(\syst)\,\GeV$.
Another recent $\mtop$ measurement obtained using ATLAS data 
taken at $\sqrt{s} = 8\,\TeV$~\cite{Aaboud:2017mae} exploits the decay
$\ttbar\to W^+ b W^- \bar{b}\to q\bar{q}'b q''\bar{q}'''\bar{b}$,
where both $W$ bosons decay into jets from charged 
quarks, $q$.  
This is a challenging measurement to make because of the large multi-jet 
background arising from various other processes of the strong interaction
described by QCD.
However, all-jets $\ttbar$ events profit from having no neutrinos 
among the decay products, so that all four-momenta can be measured 
directly.
The multi-jet background for the all-jets $\ttbar$ 
channel, while large, leads to different systematic uncertainties 
than in the case of the di-leptonic, and lepton+jets 
$\ttbar$ channels. 
Events in this analysis are required to pass the selection described in
Ref.~\cite{Aaboud:2017mae}.
To determine $\mtop$ in each $\ttbar$ event, a minimum-$\chi^2$ 
approach is adopted. 
The dominant multi-jet background in the analysis is determined 
directly from the data.
To extract the measurement of $\mtop$, a template method 
with a binned minimum-$\chi^2$ approach is employed.
For each $\ttbar$ event, two values of the ratio of three-jet 
to dijet masses, $\rtt = m_{jjj}/m_{jj}$, are obtained.
The $\rtt$ observable is chosen here because of its reduced
dependence on the JES uncertainty.
After applying a final $\chi^2$ fit, which uses matrix algebra to
include non-diagonal co-variance matrices, the measurement gives
$\mtop = 173.72\,\pm\,0.55\,(\stat)\,\pm \,1.01\,(\syst)\,\GeV$.
Figure~\ref{fig:dilepton}, right plot, shows the $\rtt$ distribution
with the corresponding total fit as well as its decomposition into
signal and the multi-jet background.
The dominant sources of systematic uncertainty in this $\mtop$
measurement come from the JES, hadronisation modelling and the
$b$JES.
Finally a combination of different $\mtop$ measurements is performed,
see Ref~\cite{Aaboud:2018zbu}. The obtained combination value is
$\mtop = 172.69\,\pm\,0.25\,({\rm stat.})\,\pm\,0.41\,({\rm syst.})\,\GeV$,
which is similar to other combined results~\cite{Aaboud:2018zbu}.

\section{Measurement of the the top-quark pole mass at $\sqrt{s} = 8$ and $7\,\TeV$}
\label{sec:polemass}

In the following the $\mtoppole$ determination from the absolute and normalised 
differential cross-sections measurement in the di-leptonic $e\mu$ channel, 
is presented~\cite{Aaboud:2017ujq}.
In this analysis one $W$ boson decays to an electron ($W\to e\nu$) 
and the other to a muon ($W\to\mu\nu$), giving rise to the 
$\ttbar\to W^+ b W^- \bar{b}\to e^{\pm}\mu^{\mp}\nu\bar{\nu}b\bar{b}$ 
final state, with opposite charged lepton, that is particularly clean. 
The analysis uses ATLAS data at $\sqrt{s}=8\,\TeV$. 
Eight differential cross-section distributions are measured, 
see Ref.~\cite{Aaboud:2017ujq}.
The measurements are made using events with an opposite-charge $e\mu$ pair 
and one or two b-jets.
The results are compared to the predictions of various NLO and LO multi-leg 
$\ttbar$ event generators, and to fixed-order perturbative QCD predictions. 
top-quark mass, and mass measurements are made by comparing the measured 
distributions to predictions from both NLO plus parton shower event 
generators and fixed-order QCD calculations.
Various techniques for extracting the top quark mass from the measured 
distributions were explored.
The most precise result was obtained from a fit of fixed-order predictions 
to all eight measured distributions simultaneously, extracting 
a value of $\mtoppole =173.2\,\pm\,0.9\,\pm\,0.8\,\pm\,1.2\,\GeV$,
where the three uncertainties come from data statistics, experimental 
systematic effects, and uncertainties in the theoretical predictions. 
A previous $\mtoppole$ measurement was extracted using $\ttbar\,+\,1$-jet events
collected with the ATLAS experiment in $\sqrt{s}=\,7\,\TeV$~\cite{Aad:2015waa}.
Events were selected by using the lepton+jets final state to identify the 
$\ttbar$ system and at least one additional jet. The $\mtoppole$ was extracted
from a measurement of the normalised differential cross section
for $\ttbar\,+\,1$-jet, as a function of the inverse of the invariant mass of
the $\ttbar\,+\,1$-jet system~\cite{Aad:2015waa}. 
The achieved measurement gave
$\mtoppole=173.7\,\pm\,1.5\,({\rm stat.})\,1.4\,({\rm syst.})^{+1.0}_{-0.5}({\rm theory})\,\GeV$,
where the theoretical uncertainties include the uncertainty due to 
missing higher orders in the perturbative NLO calculation, as well 
as uncertainties due to the PDF and the strong coupling constant, $\alpha_s$,
used in the calculations. 
The experimental uncertainty accounts for the uncertainties due to 
the imperfections in the modelling of the detector response, the 
background yield and the uncertainties arising from the signal modelling 
including hadronisation. The dominant experimental uncertainties are 
due to the JES and the initial- and final- state 
radiation modelling.
The above $\mtoppole$ results agree very well with other 
determinations of $\mtoppole$, and the available measurements
of $\mtop$~\cite{Aaboud:2018zbu}.

\end{document}